# Single Micro-hole per Pixel for Thin Ge-on-Si CMOS Image Sensor with Enhanced Sensitivity upto 1700 nm

Ekaterina Ponizovskaya-Devine, *Member, IEEE*, Ahmed S. Mayet, *Member, IEEE*, Amita Rawat, *Member, IEEE*, Ahasan Ahamed, Shih-Yuan Wang, *Life Fellow, IEEE*, Aly F. Elrefaie, *Life Fellow, IEEE*, Toshishige Yamada, *Senior Member, IEEE,* and M. Saif Islam, *Fellow, IEEE*

*Abstract—*We present a Germanium "Ge-on-Si" CMOS image sensor with backside illumination for the near-infrared (NIR) electromagnetic waves (wavelength range 300–1700 nm) detection essential for optical sensor technology. The micro-holes help to enhance the optical efficiency and extend the range to the 1.7 $\mu$m wavelength. We demonstrate an optimization for the width and depth of the nano-holes for maximal absorption in the near infrared. We show a reduction in the cross-talk by employing thin $SiO_2$ deep trench isolation in between the pixels. Finally, we show a 26–50% reduction in the device capacitance with the introduction of a hole. Such CMOS-compatible Ge-on-Si sensors will enable high-density, ultra-fast and efficient NIR imaging.

## I. INTRODUCTION

SILICON photonics has gained attention due to an increased demand for complementary metal-oxide semiconductor (CMOS) compatible near-infrared image sensors [1]–[5] for mobile imaging, digital cameras, surveillance, remote monitoring, and biometrics as against the conventional III-V based sensors. Several approaches are reported to increase the pixel density. Among them, a reduction in the pixel pitch from 2.2 µm [4] to less than 800 nm [3], [5] are a few potential solutions. Establishing a trade-off between the optical sensitivity in the pixel and the bandwidth of the imager is challenging. Absorption enhancement in a thin absorber layer by the pixel surface nanostructures or single holes per pixel [6], [7] helps attain high sensitivity without compromising the bandwidth. The reduction of pixel sizes also can have a parasitic charge exchange between neighboring pixels (crosstalk) that can be successfully reduced using deep trench isolation (DTI) [8]–[10]. However, there is a demand for wavelength near 1500-1700 nm that Si image sensors cannot provide. The sensors in near-infrared can provide the visible picture, and infrared light detection and ranging (LiDAR) scanning can provide 3D images [11]. There are researchers actively working on Ge-based sensors [11]–[15] to address the demand for 1500-1550 nm wavelength range sensitivity. It has been shown theoretically and experimentally that thin Si [16], [17], and Ge layer on Si with micro-holes can be used in photo-detectors with high quantum efficiency up to 1.7 $\mu$m wavelength.

This study proposes an imager with a thin Ge layer on Sion-insulator (SOI) substrate with a 1.12×1.12 $\mu$m$^2$ pixel size. We show the impact of incorporating a single micro-hole on the optical efficiency using Finite Time Domain (FDTD) [18] Lumerical simulations. We benchmark the proposed device performance against the existing literature to show a noticeable improvement in the power absorption at 1.7 $\mu$m wavelength. We finally show a tremendous reduction in the intrinsic device capacitance with the introduction of the micro-hole. Such Ge sensors, with an enhanced power absorption at a higher wavelength, a reduced device capacitance, high carrier mobility in Ge [19], and a CMOS-compatible fabrication process, have the potential to revolutionize short wave infrared imaging.

## II. DESIGN AND OPTICAL SIMULATION

This study considers Ge-on-Si backside-illuminated sensors layered over a signal processing circuit chip with a low-noise structure [6], [7], [9]. The simulations do not include the metal contacts. The contacts will add small optical losses that can reduce the results by a few percent. Novel IR transparent metals or highly doped poly-silicon can also be used to avoid absorption losses. The pixel array is shown schematically in Fig. 1. Each pixel is 1.12 $\mu$m wide and consists of a Si layer on $SiO_2$ and a thin layer of Ge on top of the Si layer. The thickness of Ge varied from 150 nm to 1 $\mu$m. The Si thickness is 2-2.5 $\mu$m, and the total thickness of the Ge/Si stack is 3 $\mu$m. The buried $SiO_2$ layer reduces the transmittance

Ekaterina Ponizovskaya-Devine is affiliated with W&Wsens Device, 4546 El Camino, Los Altos, CA 94022 USA, and University of California Davis, Davis, CA, 95616, USA (e-mail: e.ponizovskayadevine@ucdavis.edu).

Ahmed S. Mayet, Amita Rawat, Ahasan Ahamed, and M. Saif Islam are affiliated with University of California Davis, Davis, CA, 95616, USA.

Shih-Yuan Wang, Aly F. Elrefaie, and Toshishige Yamada are affiliated with W&Wsens Device, 4546 El Camino, Los Altos, CA 94022 USA. Toshishige Yamada is also affiliated with University of California, Santa Cruz, Santa Cruz, CA 95064 USA.

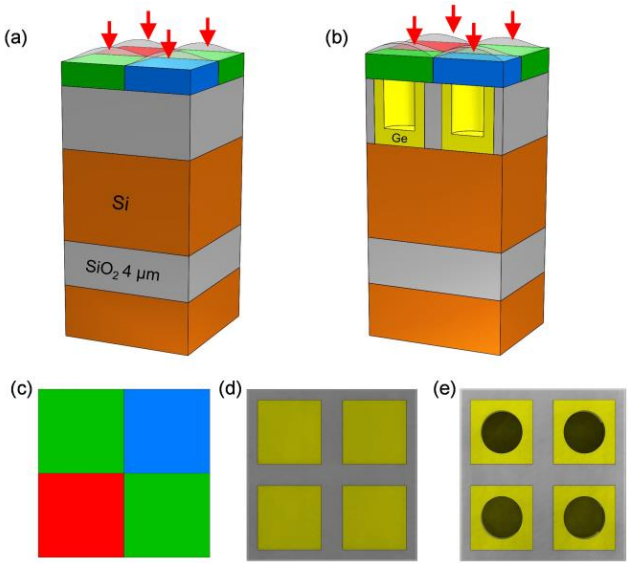

Fig. 1. The schematics of the Ge pixel: a) 3D view of the image sensor with lenses and filters, b) a section of a pixel with an air-filled cylindrical hole, c) a top view of the Bayer pattern, d) top-view of flat Ge pixel with trenches, e) top-view of holes structures in Ge pixel with trenches. Ge thickness is varied from 150 to 500 nm, with a hole diameter of 800 nm, a depth of 3/4 of Ge thickness, and Si thickness of 2.5 $\mu$m. Illumination is normal to the surface from the top (red arrows).

from the thin absorbing layer by reflecting the illuminated electromagnetic waves. A lens of radius 1 $\mu$m is used at the top of each pixel, along with a filter between the lens and the Ge layer. We model pixels in a Bayer array with commonly used color filter parameters [20]. Transmittance after the filters is shown in Fig. 2. All filters are almost transparent at 1000 nm wavelength, and we assume they have the same characteristics in infrared.

In the 3D simulation setup, the transmittance associated with 900 nm thickness is used for the pigment filters [1], [21]–[23], and the parameters are listed in Table I. The estimated optical efficiency (OE) 1 $\mu$m and 500 nm Ge layer are shown in Fig. 3(a) and (b), respectively. The transmission profile of the flat device (no photon-trapping structure on the surface) for infrared is shown with a black line. We calculate the OE, which represents the absorption of light by the active layer of each pixel. The quantum efficiency (QE) on the other hand is the photocurrent of the device normalized to the light intensity and is less than the OE due to recombination and other losses. Still, the higher OE translates to a higher QE, and optimizing the light-trapping quality would increase QE. The maximum optical efficiency for the blue, red, and green filters determined by the filter transmittance is $OE_{Blue}$ = 75% at

440 nm, $OE_{Green}$ = 80% at 550 nm, and $OE_{Red}$ = 80% at 650 nm (Fig. 3(a-b)). In contrast, the filters are shown to be transparent in the near-infrared (Fig. 2). The Ge optical absorption is very weak at 1500 nm, and light-trapping strategies are required to enhance the absorption, leading to higher optical efficiency.

To increase the OE, we incorporate a cylindrical hole at the center of each pixel (Fig. 1(e)). The diameter of the hole is 800 nm. There are trenches at the edges of each pixel with a width

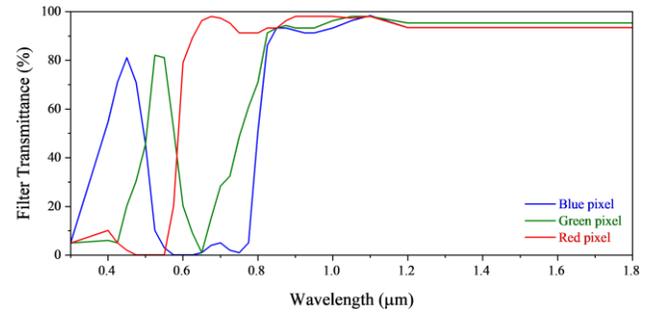

Fig. 2. Transmittance through the filters.

of 150 nm to the crosstalk. The depth of the holes and trenches is the same as the thickness of the Ge layer. Micro-hole arrays were shown to enhance the optical absorption of Si for the 800-1000 nm wavelength range [16] and the Ge for 1400-1800 nm wavelength [17]. The study [6], [7] shows that a single hole per pixel provides a better increase in quantum efficiency. The optimal micro hole size is comparable to the small-sized pixels. The deep trench structure with the Si-SiO$_2$ interface is used as a barrier against electron diffusion and to prevent crosstalk. The micro-structure redirects the normal incident light into lateral directions parallel to the surface plane and helps increase the absorption and reduce reflection [17], [24]. We used the FDTD method provided by the Lumerical software to solve Maxwell's curl equations numerically for the unit cell that consists of the Bayer array (Fig. 1). Bloch boundary conditions in the *XY* plane and Perfectly Matched Layer (PML) in the direction normal (*Z*) to the surface were considered. The CMOS image sensor model includes lenses, red, green, and blue filters with a thickness of 900 nm, antireflection coating, and a 150 nm to 1 $\mu$m thick Ge-on-Si and SOI substrate micro-lens with a radius of 1 $\mu$m and a thickness of 500 nm. (as shown in Fig. 1(b))

The total absorbed power ($P_{abs}$) is simulated around the cells.

TABLE I
FILTERS PARAMETERS [1], [21]–[23]

| $\lambda$ (nm) | Blue | | Green | | Red | |
|---|---|---|---|---|---|---|
| | n | k | n | k | n | k |
| 300 | 1.55 | 0.055 | 1.62 | 0.400 | 1.54 | 0.4150 |
| 400 | 1.54 | 0.070 | 1.60 | 0.500 | 1.54 | 0.3250 |
| 500 | 1.54 | 0.215 | 1.58 | 0.405 | 1.53 | 0.1550 |
| 600 | 1.54 | 0.800 | 1.57 | 0.050 | 1.53 | 0.0480 |
| 700 | 1.53 | 0.450 | 1.57 | 0.110 | 1.52 | 0.0330 |
| 800 | 1.53 | 0.455 | 1.57 | 0.105 | 1.52 | 0.3750 |
| 900 | 1.52 | 0.465 | 1.56 | 0.090 | 1.52 | 0.0185 |
| >1000 | 1.52 | 0.390 | 1.56 | 0.055 | 1.52 | 0.0150 |

The absorbed power in the pixel with volume $V$ is calculated using the electric ($E$) and magnetic ($H$) components of the optical field using eq. (1).

$$P_{abs} = \int_V \frac{1}{2} \nabla \cdot Re(E \times H^*) \, dv$$
$$= \oint_S \frac{1}{2} \left[ Re(E \times H^*) \cdot \vec{n} \right] ds \quad (1)$$

Where $S$ is the surface surrounding the volume $V$, which includes filters and the absorbing layer, and $\vec{n}$ is the unit vector normal to the surface $S$. The OE is the ratio of the absorbed power ($P_{abs}$) on the surface $S$ (as determined from eq. (1)) to the incident power ($P_{in}$) is given by eq. (2).

$$OE = \frac{P_{abs}}{P_{in}} \quad (2)$$

$P_{in}$ is the total power of the incident light calculated above the lenses and filters. We assume that the quantum efficiency is proportional to the optical efficiency [13]. The filter's response spectrum was calculated from the refractive index presented in Table I. For the wavelength higher than 1000 nm the simulations assumed the same filter parameters as for 1000 nm. It means that the filters are almost transparent in infrared. We have studied the influence of Ge thickness on optical efficiency. We reduced the Ge thickness from 1000 nm to as low as 150 nm. As is expected, the optical efficiency is reduced with the thickness. However, the reduction is almost negligible up to 250 nm and starts reducing for even thinner Ge. Although a flat pixel of thickness 150 nm produces minimal optical efficiency in infrared, the single cylindrical hole increases the efficiency to about 40%, which is still useful for sensing infrared. A Ge layer, thinner than 150 nm, will not be practical.

The study [13] of an image sensor with Ge showed significant quantum efficiency at 400 nm Ge. Our study shows that using a single hole per pixel is a way to significantly improve the quantum efficiency for an even thinner Ge layer for an extended range of wavelength—up to 1700 nm.

The optical efficiency in the visible range is high due to the high absorption of Ge for the visible wavelengths. For extremely thin Ge at 150 nm, the optical efficiency is reduced from 80% to 60%, remaining high enough. A thin absorbing layer could be interesting from the fabrication point of view because it will not cause wafer bending or elongated threading dislocations due to lattice mismatch with the Si substrate [25]. Also, it is interesting because a thinner absorbing layer can provide a faster performance due to the reduced conduction path between the two electrodes. The hole depth is ¾ of the Ge thickness filled with SiO$_2$. The OE in the visible range is shown in Fig. 3(c) for 150 nm, which is lower than 500 nm. The infrared OE is demonstrated in Fig. 3(d) for 350 nm and in Fig. 3(e) for 150 nm (red line) compared against a flat device (black line) and a device with 500 nm Ge thickness (blue line). As one can see, the 350 nm with cylindrical holes device produces almost the same OE as that of 500 nm. However, the 150 nm devices show a significantly reduced OE. The optical efficiency in infrared for all the Ge thicknesses is significantly higher than that for the flat device.

An increase in the crosstalk between pixels could be an undesired side-effect of this increased OE. The crosstalk is desired to be reduced using deep trenches isolators (DTI) [7]. We have presented the crosstalk index between each

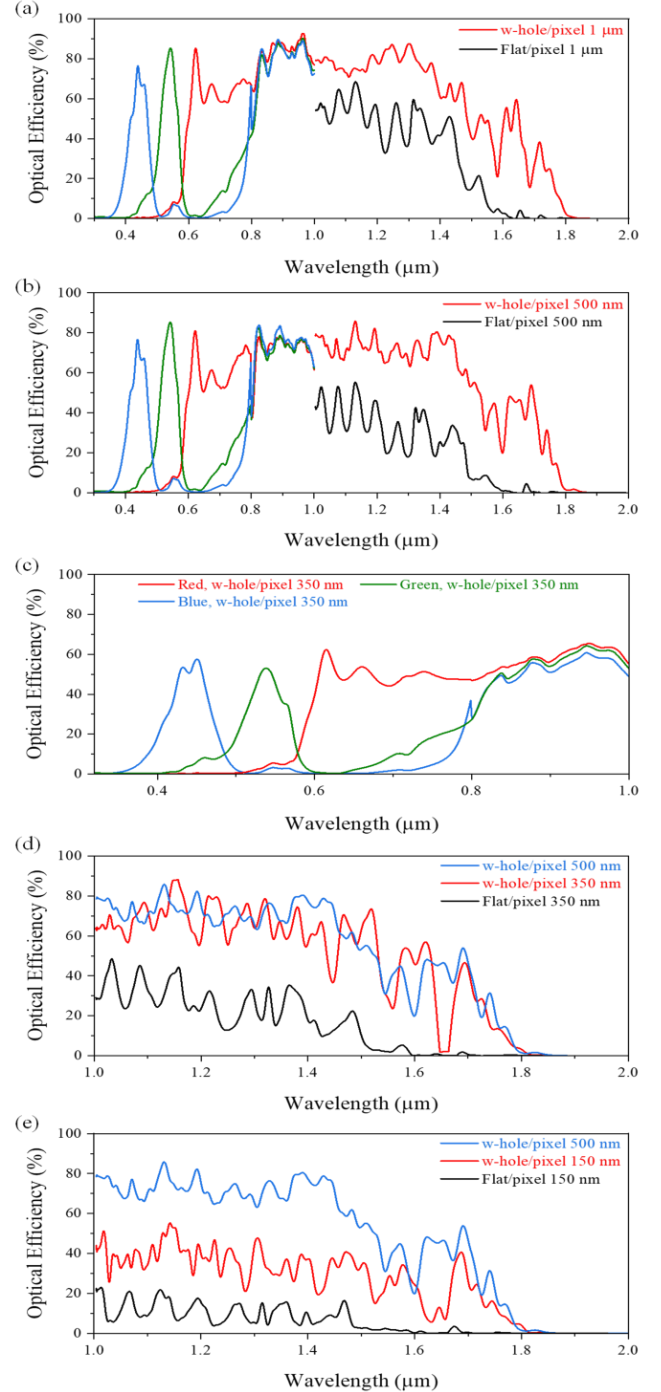

Fig. 3. Optical efficiency of the red, blue, and green pixels marked with the respective colors, and the optical efficiency for flat pixels marked with the black color. Optical efficiency trends with cylindrical holes sensors of (a) 1 μm and (b) 500 nm Ge thicknesses for a large wavelength range; (c) an optical efficiency of the red, blue, and green pixels in the visible spectrum for 350 nm Ge thickness; optical efficiency trends for the with cylindrical hole sensors for Ge thicknesses (d) 350 nm and (e) 150 nm in infrared wavelength range, and are compared against a cylindrical hole Ge sensor of Ge thickness 500 nm. The hole depth is 3/4 of the Ge thickness.

TABLE II

CROSSTALK BETWEEN PIXELS WITH DIFFERENT GE THICKNESSES IN THE VISIBLE RANGE

| Pixel Color | Crosstalk Index Ge | | |
|---|---|---|---|
| | 500 nm | 350 nm | 150 nm |
| Green-red | 27 | 26 | 20 |
| Red-green | 16 | 15 | 9 |
| Red-blue | 82 | 82 | 50 |
| Blue-red | 85 | 82 | 50 |
| Green-blue | 31 | 29 | 20 |
| Blue-green | 5 | 4 | 3.5 |

pair of colors, e.g., the red-green crosstalk, i.e., the response of the green pixel at the red wavelength, in our previous work [7]. This definition implies that a higher value of the index of crosstalk corresponds to higher contrast and lower crosstalk. The trenches could be optimized to reduce the crosstalk by varying depths and thicknesses. We have utilized the previous results for trench depth optimization, as shown in Table II. It is defined as the OE value for the maximum for the corresponding color to the OE value of the neighbor pixel at the same wavelength. As we can see, the crosstalk is negligible for 500 nm Ge. The crosstalk is slightly increasing with the decrease in the Ge thickness and the depth of the holes. However, even for 150 nm, we have a relatively small crosstalk index compared to what was previously reported in the literature [24]. In the design, we used the DTI technique to reduce crosstalk, which is effective even for 150 nm Ge. We have compared the performance of the proposed Ge sensor with the state-of-the-art literature in Table III. We report an increased OE in the 1.70 $\mu$m wavelength range.

TABLE III

DEVICE BENCH-MARKING AGAINST STATE-OF-THE-ART LITERATURE

| Ref. (Year) | Sub. | Active region ($\mu$m) | Pixel pitch ($\mu$m) | QE[1]/OE[2] (%) | | |
|---|---|---|---|---|---|---|
| | | | | @1310 nm | @1550 nm | @1700 nm |
| [12] (2009) | Ge-on-Si | 2-5.6 | 7-10 | 60[1] | 10[1] | - |
| [14] (2011) | Ge-on-Si | 1.5 | 120 | 37[1] | 22[1] | - |
| [15] (2019) | Ge-on-Si | 0.6 | 30[1] | 38[1] | 16[1] | <1 |
| [13] (2022) | Ge-on-Si | 0.4 | 25[1] | 34[1] | 4[1] | <1 |
| This work (Simulation) | Ge-on-SOI | 0.15 | 1.12 | 35[2] | 30[2] | 37[2] |
| | Ge-on-SOI | 0.35 | 1.12 | 76[2] | 42[2] | 47[2] |
| | Ge-on-SOI | 0.50 | 1.12 | 68[2] | 40[2] | 51[2] |

## III. ELECTRICAL SIMULATION

To understand the electrical performance and compare the impact of incorporating the holes in the sensor, we have performed ATLAS Silvaco [26] device simulations. We have simulated a device structure with a 200 nm top p-Ge layer with doping 5×10$^{18}$ cm$^{-3}$ followed by a 300 nm p-Ge layer with 1×10$^{15}$ cm$^{-3}$ doping (intrinsic region) stacked over a 2.0 $\mu$m Si layer (doping 5×10$^{18}$ cm$^{-3}$). The metal contacts are at the top and the bottom of the device as shown in Fig. 4(a).

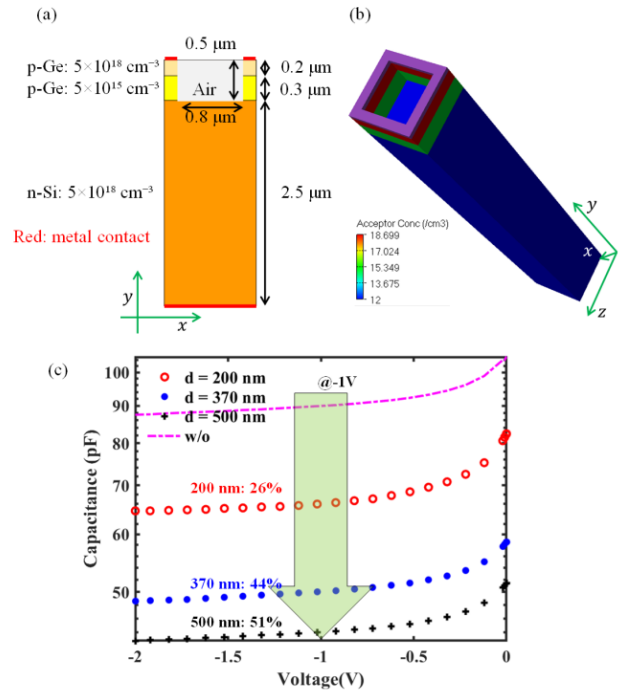

Fig. 4. (a) Schematic of the Ge image sensor simulated on ATLAS Silvaco; (b) 3D with-hole Ge sensor showing the doping concentration of the p-Ge and i-Ge layer; (c) capacitance versus voltage profile extracted from simulation.

A 3D Ge-sensor structure showcasing the acceptor doping profile is shown in Fig. 4(b). We have simulated four different structures as follows: 1) without-hole device; 2) 200 nm hole depth; 3) 370 nm hole depth; 4) 500 nm hole depth. The hole width is fixed at 800 nm. We have used Shockley Read Hall, Auger, and field-dependent mobility models to incorporate mobility accurately. We have utilized the Poisson and drift-diffusion models for carrier transport. Due to the high doping, we have used the Fermi probability distribution function instead of Boltzmann's approximation. Further details of the simulation is added in the appendix. To study the impact of the hole on the capacitance of the device, we performed a capacitance-voltage (C-V) simulation on the devices. The obtained C-V profile is shown in Fig. 4(c). We have compared the C among with-hole and without-hole devices. The capacitance values show a significant reduction with the introduction of the hole as opposed to the flat device. We see a gradual decrease from 26% to 50% with the hole depth increasing from 200 nm to 500 nm. This reduction can be attributed to the reduced effective volume of the with-hole devices. Further, we have shown the impact of introducing the hole on the electrostatic (ES) potential distribution and current density using a 2D cross-section contours as shown in Fig. 5. In contact. With the introduction of the hole in Fig.5(a-ii), the potential distribution in the i-Ge region remains unchanged. In Fig. 5(b), we have compared the current density (CD) profile for w/o hole and w/ hole devices. We show that the high CD region near the contact in the w/o hole device (Fig. 5(b-i)) is shifted toward the i-Ge/n-Si junction (Fig. 5(b-ii)). By introducing the hole, we have redirected the lateral current conduction through p-Ge region toward the i-Ge region. This redirection is essential for cultivating the generated $e^-$-$h^+$ pair. The main motivation for electrical simulation is to estimate the intrinsic device capacitance with the introduction of the hole, therefore, we have not considered the impact of surface states that leads to a high off-state current in this work. We observe a noticeable decrease in the capacitance when incorporating the holes into the Ge sensors.

Fig. 5(a-i), we show the ES potential distribution in w/o hole device mainly drops in the i-Ge region, i.e., it creates a high electric field required to accelerate the generated electron-hole ($e^-$-$h^+$) pair toward the

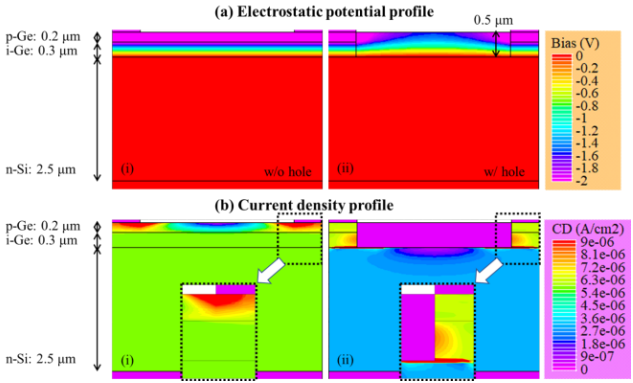

Fig. 5. Contour plots showing a comparison of (a) electrostatic potential, and (b) current density profiles for w/o hole versus w/ hole devices.

## IV. CONCLUSION

We have shown that Ge CMOS image sensors with cylindrical holes exhibit a high optical efficiency of up to 1.7 $\mu m$ wavelength. The crosstalk increased due to holes, but it was reduced back to a normal level with the implementation of trenches between individual pixels. The trenches with 150 nm width and a depth equal to the depth of Si provide negligible crosstalk. The simulations show that the 500 nm Ge thickness increases OE in infrared while 1 $\mu m$ Ge thickness doesn't produce much additional improvement. At the same time, smaller thicknesses provide faster device performance. It means that the thickness of 500 nm is preferable and still gives maximal efficiency. The Ge thickness could be reduced to 150 nm with cylindrical holes with 40% optical efficiency in infrared. The minimal thickness when the efficiency is still sufficient for infrared sensing. Ge CMOS sensors can provide much better optical efficiency in infrared even for the very thin thickness of Ge. Along with the gain on the optical absorption fronts, we report a significant reduction of 26–50% in the device capacitance with the introduction of the micro-hole.

## APPENDIX: SILVACO SIMULATION DETAILS

The electrical simulations are performed on ALTAS Silvaco technology computer-aided design software. To optimize the performance and the computation time, we have used a nonuniform doping profile. In the following subsections, we have described the device physics models used in the simulation. *A. Field-dependent electron (hole) mobility*

$$\mu_{n(p)} = \mu_{n(p)0} \left[ \frac{1}{1 + \left(\frac{\mu_{n(p)}E}{V_{sat,n(p)}}\right)^{\beta_{n(p)}}} \right]^{1/\beta_{n(p)}}$$

where $\mu_{n(p)0}$ is low field electron (hole) mobility, $E$ is the parallel electric field, $V_{sat,n(p)}$ is saturation velocity, and $\beta_{n(p)}$ is user-defined calibration parameter.
[$\beta_n$ = 2; $\beta_p$ = 1; $\mu_{n(p)0-Si}$ = 1500 (450) $cm^2/Vs$; $\mu_{n(p)0-Ge}$ = 3900 (1900) $cm^2/Vs$; $n_{i-Si}$ = 1 × $10^{10}$ $cm^{-3}$; $n_{i-Ge}$ = 2 × $10^{13}$ $cm^{-3}$].

*B. Recombination and Generation*
*1) Shockley-Read-Hall recombination:*

$$R_{SRH} = \frac{pn - n_i^2}{\tau_{p0}\left(n + \frac{n_i e^{E_F - E_T}}{kT}\right) + \tau_{n0}\left(p + \frac{n_i e^{-E_F + E_T}}{kT}\right)}$$

$n$ and $p$ are electron and hole carrier concentrations, $n_i$ is intrinsic carrier concentration, $E_F$ and $E_T$ are Fermi energy and trap energy levels, $\tau_{n0}$ and $\tau_{p0}$ are the electron and the hole lifetimes and are assumed to be same for Si and Ge.[$E_F - E_T$ = 0 $eV$ ; $\tau_{n0} = \tau_{p0}$ = 1 × $10^{-7}$ $s$; $T$ = 300$K$].
*2) Auger Recombination [27]:*

$$R_{Auger} = C_n(pn^2 - nn_i^2) + C_p(np^2 - pn_i^2)$$

$C_n$ and $C_p$ are Auger recombination coefficients [27]. [$C_n$ = 2.8 × $10^{-31}$ $cm^6/s$; $C_p$ = 9.9 × $10^{-32}$ $cm^6/s$]

With this device physics, and Fermi-Dirac electron probability distribution, the ATLAS solved Poisson and continuity equations self-consistently to produce the electrical characteristics of the device under test.

## ACKNOWLEDGMENT

This work was supported in part by the Dean's Collaborative Research Award (DECOR) of UC Davis College of Engineering, the S. P. Wang and S. Y. Wang Partnership, Los Altos, CA.

## REFERENCES

[1] C.-F. Han, J.-M. Chiou, and J.-F. Lin, "Deep trench isolation and inverted pyramid array structures used to enhance optical efficiency of photodiode in cmos image sensor via simulations," *Sensors*, vol. 20, no. 11, p. 3062, 2020.

[2] I. Oshiyama, S. Yokogawa, H. Ikeda, Y. Ebiko, T. Hirano, S. Saito, T. Oinoue, Y. Hagimoto, and H. Iwamoto, "Near-infrared sensitivity enhancement of a back-illuminated complementary metal oxide semiconductor image sensor with a pyramid surface for diffraction structure," in *2017 IEEE International Electron Devices Meeting (IEDM)*, pp. 16–4, IEEE, 2017.

[3] S. Yokogawa, I. Oshiyama, H. Ikeda, Y. Ebiko, T. Hirano, S. Saito, T. Oinoue, Y. Hagimoto, and H. Iwamoto, "Ir sensitivity enhancement of cmos image sensor with diffractive light trapping pixels," *Scientific reports*, vol. 7, no. 1, pp. 1–9, 2017.

[4] R. Fontaine, "The state-of-the-art of smartphone imagers," in *International Image Sensor Workshop*, p. R01, 2019.

[5] J. Ahn, C.-R. Moon, B. Kim, K. Lee, Y. Kim, M. Lim, W. Lee, H. Park, K. Moon, J. Yoo, *et al.*, "Advanced image sensor technology for pixel scaling down toward 1.0 $\mu$m," in *2008 IEEE International Electron Devices Meeting*, pp. 1–4, IEEE, 2008.

[6] E. P. Devine, W. Qarony, A. Ahamed, A. S. Mayet, S. Ghandiparsi, C. Bartolo-Perez, A. F. Elrefaie, T. Yamada, S.-Y. Wang, and M. S. Islam, "Single microhole per pixel in cmos image sensors with enhanced optical sensitivity in near-infrared," *IEEE Sensors Journal*, vol. 21, no. 9, pp. 10556–10562, 2021.

[7] E. P. Devine, A. Ahamed, A. S. Mayet, S. Ghandiparsi, C. BartoloPerez, L. McPhillips, A. F. Elrefaie, T. Yamada, S.-Y. Wang, and M. S. Islam, "Optimization of cmos image sensors with single photon-trapping hole per pixel for enhanced sensitivity in near-infrared," *arXiv preprint arXiv:2110.00206*, 2021.

[8] B. J. Park, J. Jung, C.-R. Moon, S. H. Hwang, Y. W. Lee, D. W. Kim, K. H. Paik, J. R. Yoo, D. H. Lee, and K. Kim, "Deep trench isolation for crosstalk suppression in active pixel sensors with 1.7 $\mu$m pixel pitch," *Japanese journal of applied physics*, vol. 46, no. 4S, p. 2454, 2007.

[9] A. Tournier, F. Leverd, L. Favennec, C. Perrot, L. Pinzelli, M. Gatefait, N. Cherault, D. Jeanjean, J. Carrere, F. Hirigoyen, *et al.*, "Pixel-to-pixel isolation by deep trench technology: Application to cmos image sensor," in *Proc. Int. Image Sensor Workshop*, pp. 12–15, 2011.

[10] "Imaging and sensing technology | imaging and sensing technology | sony semiconductor solutions group. https://www.sonysemicon.co.jp/e/technology/imaging-sensing/ (accessed nov. 22, 2020).,"

[11] A. Sammak, M. Aminian, L. Qi, E. Charbon, and L. K. Nanver, "A 270× 1 ge-on-si photodetector array for sensitive infrared imaging," in *Optical Sensing and Detection III*, vol. 9141, pp. 16–22, SPIE, 2014.

[12] B. Ackland, C. Rafferty, C. King, I. Aberg, J. O'Neill, T. Sriram, A. Lattes, C. Godek, and S. Pappas, "A monolithic ge-on-si cmos imager for short wave infrared," in *International Image Sensor Workshop*, 2009.

[13] M. Oehme, M. Kaschel, S. Epple, M. Wanitzek, Z. Yu, D. Schwarz, A.C. Kollner, J. N. Burghartz, and J. Schulze, "Backside illuminated "ge-¨ on-si" nir camera," *IEEE Sensors Journal*, vol. 21, no. 17, pp. 18696–18705, 2021.

[14] R. Kaufmann, G. Isella, A. Sanchez-Amores, S. Neukom, A. Neels, L. Neumann, A. Brenzikofer, A. Dommann, C. Urban, and H. von Kanel,¨ "Near infrared image sensor with integrated germanium photodiodes.," *Journal of Applied Physics*, vol. 110, no. 2, p. 023107, 2011.

[15] A.-C. Kollner, Z. Yu, M. Oehme, J. Anders, M. Kaschel, J. Schulze,¨ and J. N. Burghartz, "A 2x2 pixel array camera based on a backside illuminated Ge-on-Si photodetector," in *2019 IEEE SENSORS*, pp. 1–4, IEEE, 2019.

[16] Y. Gao, H. Cansizoglu, K. G. Polat, S. Ghandiparsi, A. Kaya, H. H. Mamtaz, A. S. Mayet, Y. Wang, X. Zhang, T. Yamada, *et al.*, "Photontrapping microstructures enable high-speed high-efficiency silicon photodiodes," *Nature Photonics*, vol. 11, no. 5, pp. 301–308, 2017.

[17] H. Cansizoglu, C. Bartolo-Perez, Y. Gao, E. P. Devine, S. Ghandiparsi, K. G. Polat, H. H. Mamtaz, T. Yamada, A. F. Elrefaie, S.-Y. Wang, *et al.*, "Surface-illuminated photon-trapping high-speed ge-on-si photodiodes with improved efficiency up to 1700 nm," *Photonics Research*, vol. 6, no. 7, pp. 734–742, 2018.

[18] ""high-performance photonic simulation software - lumerical." https://www.lumerical.com/ (accessed nov. 22, 2020).,"

[19] F. Schaffler, "High-mobility si and ge structures,"¨ *Semiconductor Science and Technology*, vol. 12, no. 12, p. 1515, 1997.

[20] C. Park and M. G. Kang, "Color restoration of rgbn multispectral filter array sensor images based on spectral decomposition," *Sensors*, vol. 16, no. 5, p. 719, 2016.

[21] A. Taflove, S. C. Hagness, and M. Piket-May, "Computational electromagnetics: the finite-difference time-domain method," *The Electrical Engineering Handbook*, vol. 3, pp. 629–670, 2005.

[22] J. Berzins, S. Fasold, T. Pertsch, S. M. Bˆ aumer, and F. Setzpfandt,¨ "Submicrometer nanostructure-based RGB filters for CMOS image sensors," *ACS Photonics*, vol. 6, no. 4, pp. 1018–1025, 2019.

[23] H. R. Myler, A. R. Weeks, and L. Voicu, "RGB color enhancement using homomorphic filtering," in *Image and Video Processing III*, vol. 2421, pp. 43–50, SPIE, 1995.

[24] Y. Horie, S. Han, J.-Y. Lee, J. Kim, Y. Kim, A. Arbabi, C. Shin, L. Shi, E. Arbabi, S. M. Kamali, *et al.*, "Visible wavelength color filters using dielectric subwavelength gratings for backside-illuminated cmos image sensor technologies," *Nano Letters*, vol. 17, no. 5, pp. 3159–3164, 2017.

[25] A. Abedin, *Germanium layer transfer and device fabrication for monolithic 3D integration*. PhD thesis, KTH Royal Institute of Technology, 2021.

[26] "Atlas: https://silvaco.com/tcad/,"

[27] A. P. Kirk, *Solar photovoltaic cells: photons to electricity*. Academic Press, 2014.